\documentclass[aps,twocolumn,a4paper,showpacs,nofootinbib,superscriptaddress]{revtex4}
\usepackage[dvips]{graphicx}
\usepackage{amssymb}

\newcommand{\sgn}{\mathop{\mathrm{sgn}}\nolimits}

\begin{document}

\title{Quantum capacitance and compressibility of graphene:\\
The role of Coulomb interactions}
\author{Yu.E. Lozovik}%
\email{lozovik@isan.troitsk.ru}
\affiliation{Institute for Spectroscopy, Russian Academy of Sciences, 142190 Troitsk, Moscow, Russia}%
\affiliation{MIEM at National Research University HSE, 109028 Moscow, Russia}
\affiliation{All-Russia Research Institute of Automatics, 127055 Moscow, Russia}%
\affiliation{National Research Nuclear University MEPHI, 115409 Moscow, Russia}%
\author{A.A. Sokolik}%
\affiliation{Institute for Spectroscopy, Russian Academy of Sciences, 142190 Troitsk, Moscow, Russia}%
\affiliation{MIEM at National Research University HSE, 109028 Moscow, Russia}%
\author{A.D. Zabolotskiy}%
\affiliation{Institute for Spectroscopy, Russian Academy of Sciences, 142190 Troitsk, Moscow, Russia}%
\affiliation{All-Russia Research Institute of Automatics, 127055 Moscow, Russia}%
\begin{abstract}
Many-body effects on quantum capacitance, compressibility, renormalized Fermi velocity, kinetic and interaction
energies of massless Dirac electrons in graphene, induced by Coulomb interactions, are analyzed theoretically in the
first-order, Hartree-Fock and random phase approximations. Recent experimental data on quantum capacitance and
renormalized Fermi velocity are analyzed and compared with the theory. The bare Fermi velocity and the effective
dielectric constants are obtained from the experimental data. A combined effect of Coulomb interactions and Gaussian
fluctuations of disorder potential is considered.
\end{abstract}


\maketitle

\section{Introduction}

Discovery of graphene, a two-dimensional carbon material with effectively massless electrons, stimulated new
fundamental and applied studies in solid state physics \cite{CastroNeto,Abergel_Review,Kotov1}. In recent years,
considerable attention has been attracted to the problem of compressibility and quantum capacitance of graphene, which
is connected both with fundamental physics of the Coulomb-interacting gas of massless electrons and with possible
applications of graphene in electronics and energy storage technologies.

In the early experiments \cite{Martin} on graphene electron compressibility, results consistent with the model of a
non-interacting Dirac electron gas were reported. The linear dispersion and chirality of graphene electrons were
proposed as possible causes of the apparent absence of electron interaction signatures \cite{Abergel1}.

The recent experiments \cite{Yu,Chen2,Kretinin} on measuring electron compressibility or quantum capacitance in
high-quality graphene samples revealed signatures of electron interactions, in consistency with the many-body
calculations \cite{Hwang,Li_Hwang,Barlas,Asgari2,Sheehy} of these quantities. The observed interaction-induced effects
are closely related to the logarithmic renormalization of the electron Fermi velocity in graphene in the vicinity of
the charge neutrality point (CNP), which was observed by different experimental groups
\cite{Li,Elias,Siegel,Luican,Chae} and considered in theoretical literature (see \cite{Gonzalez1}, reviews
\cite{CastroNeto,Abergel_Review,Kotov1} and literature cited in \cite{Barnes}).

A random potential, arising in real graphene samples due to charged impurities and corrugations, manifests itself in
formation of electron-hole puddles \cite{Martin,Chae,Martin2,Hajaj,Zhang} and qualitatively changes graphene physics at
low carrier densities near CNP. Disorder has been proposed as a source of the observed nonvanishing compressibility and
quantum capacitance of graphene at CNP \cite{Droscher,Chen1,Xia1,Xia2,Chen2,Ponomarenko,Hajaj,Xu}. To describe the
experimentally measured dependencies of compressibility and quantum capacitance on electron density the model of
Gaussian fluctuations of the disorder potential was successfully used \cite{Hajaj,Kliros,Zebrev,Abergel2,Li2,Xu}. The
random phase approximation with a polarizability, modified by disorder, was used to calculate the compressibility in
\cite{Asgari2}.

In the present article, we perform a theoretical study of quantum capacitance and related properties of graphene in
presence of Coulomb interactions in the first-order approximation (FOA), Hartree-Fock approximation (HFA) and random
phase approximation (RPA). In order to obtain the bare Fermi velocity $v_\mathrm{F}$, we analyze the recent
experimental data on quantum capacitance and renormalized Fermi velocity \cite{Yu,Chen2,Kretinin,Chae}. Influence of
Coulomb interactions on quantum capacitance and renormalized Fermi velocity (see Sec. \ref{sect2} for its definition)
as well as kinetic and interaction energies of an electron gas in graphene are studied in FOA, HFA and RPA. A combined
effect of Coulomb interaction and disorder on these quantities is studied within the model of Gaussian electrostatic
potential fluctuations.

We show that both HFA and RPA are in close agreement with the experiments at
$v_\mathrm{F}\approx0.9\times10^6\,\mbox{m/s}$, although HFA requires much larger effective dielectric constants of
surrounding media to simulate the screening, lacking in this approximation. The influence of Coulomb interactions on
the properties of the electron gas has two major features: exchange effects push the Fermi velocity to higher values
and the quantum capacitance to lower values; correlation effects partly compensate the exchange ones. The renormalized
Fermi velocity increases up to 50\,\% at the lowest achievable densities near CNP and by 10--20\,\% away from CNP. The
quantum capacitance is typically reduced by 10--15\,\%, although it can be described within the non-interacting model
with $v_\mathrm{F}\approx1.1\times10^6\,\mbox{m/s}$. In presence of disorder, a nonzero quantum capacitance appears at
CNP, in agreement with the experiments, whereas the renormalized Fermi velocity turns out to be suppressed near CNP.

The article is organized as follows. Is Sec.~\ref{sect1} we present theoretical models used to calculate the
characteristics of the electron gas. In Sec.~\ref{sect2} we perform an analysis of experimental data. Many-body effects
of Coulomb interactions on the properties of the electron gas in graphene are studied in Sec.~\ref{sect3}. Influence of
disorder is considered in Sec.~\ref{sect4}, and Sec.~\ref{concl} concludes the article.

\section{Theoretical models}\label{sect1}

We start with a description of the electron gas in graphene in terms of a grand canonical ensemble when the temperature
$T$, chemical potential $\mu$ and area of the system $S$ are the controlling parameters. Physically this corresponds to
a flake of graphene, brought in a contact with a conductor, specifying $\mu$. Under these conditions, the system tends
to an equilibrium, where the thermodynamic potential $\Omega=E-T\mathfrak{S}-\mu N$ attains a minimum ($E$ and
$\mathfrak{S}$ are the internal energy and entropy of the electron gas, $N$ is the mean number of electrons in the
system). The electron surface density, or concentration, $n=N/S$, is given by
\begin{eqnarray}
n=-\frac1S\frac{\partial\Omega}{\partial\mu}.\label{n}
\end{eqnarray}
The compressibility $\kappa$ and the quantum capacitance per unit area $C_\mathrm{Q}$ can be calculated as:
\begin{eqnarray}
\kappa=\frac1{n^2}\frac{dn}{d\mu},\qquad C_\mathrm{Q}=e^2\frac{dn}{d\mu}\label{compressibility}
\end{eqnarray}
(sometimes merely $d\mu/dn$ is referred to as the inverse compressibility \cite{Hwang,Chen2}). The quantum
$C_\mathrm{Q}$ and classical $C_\mathrm{C}$ capacitances form the total capacitance $C_\mathrm{tot}$ as
$C_\mathrm{tot}^{-1}=C_\mathrm{Q}^{-1}+C_\mathrm{C}^{-1}$, thus the smaller of them dominates. In particular,
$C_\mathrm{tot}$ acquires a significant quantum correction when $C_\mathrm{Q}\ll C_\mathrm{C}$ (see insets in
Fig.~\ref{Fig6}).

In a noninteracting system, the thermodynamic potential $\Omega_0$ in the $T\rightarrow0$ limit is:
\begin{eqnarray}
\Omega_0=g\sum_{\mathbf{p}\gamma}(\epsilon_{\mathbf{p}\gamma}-\mu)f(\epsilon_{\mathbf{p}\gamma}),\label{Omega0}
\end{eqnarray}
where $\epsilon_{\mathbf{p}\gamma}=\gamma v_\mathrm{F}|\mathbf{p}|$ is the one-particle energy of an electron in
graphene with the momentum $\mathbf{p}$ in a conduction or valence band at $\gamma=\pm1$ respectively; $v_\mathrm{F}$
is the bare Fermi velocity; $g=4$ is the degeneracy factor over spin and valleys; and
$f(\epsilon)=\Theta(\mu-\epsilon)$ is the occupation number for a state with the energy $\epsilon$ at $T\rightarrow0$,
where $\Theta(x)$ is the unit step function.

\begin{figure}[t]
\begin{center}
\resizebox{0.6\columnwidth}{!}{\includegraphics{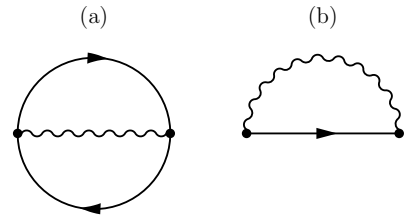}}
\end{center}
\caption{\label{Fig1}(a) The first-order exchange diagram for $\delta\Omega$ (\ref{Omega_1st}). (b) The first-order
exchange diagram for the self-energy (\ref{Sigma_1st}).}
\end{figure}

The electron density in the noninteracting graphene is determined through (\ref{n}) and (\ref{Omega0}) as:
\begin{eqnarray}
n_0(\mu)=\sgn(\mu)\frac{g\mu^2}{4\pi v_\mathrm{F}^2}\label{n0}
\end{eqnarray}
($\hbar\equiv1$). Here $\mu$ and $n$ are counted, respectively, from CNP and from the background electron density of
the filled valence band; thus $n$ is positive or negative in, respectively, electron- or hole-doped graphene.

Change of the thermodynamic potential $\delta\Omega=\Omega-\Omega_0$ when interactions are switched on can be
calculated as a sum of closed connected diagrams \cite{Abrikosov,Mahan}.

The simplest approximation to calculate $\Omega$ is the first-order approximation, where we take into account only the
first-order exchange diagram, shown in Fig.~\ref{Fig1}(a). The resulting first-order correction to the thermodynamic
potential is:
\begin{eqnarray}
\delta\Omega_1=-\frac{g}{2S}\sum_{\mathbf{p}\mathbf{p}'\gamma\gamma'}V_{\mathbf{p}-\mathbf{p}'}
F_{\mathbf{p}\gamma\mathbf{p}'\gamma'}f(\epsilon_{\mathbf{p}\gamma})f(\epsilon_{\mathbf{p}'\gamma'}),\label{Omega_1st}
\end{eqnarray}
where $V_\mathbf{q}=2\pi e^2/\varepsilon|\mathbf{q}|$ is the Coulomb potential, $\varepsilon$ is the effective
dielectric permittivity of a surrounding medium,
$F_{\mathbf{p}\gamma\mathbf{p}'\gamma'}=[1+\gamma\gamma'\cos(\mathbf{p}\hat{\;}\mathbf{p}')]/2$ is the angular factor,
accounting for an overlap of two-component spinor parts of electron wave functions.

From (\ref{n}) and (\ref{Omega_1st}) we get the first-order correction $\delta n_1=n-n_0$ to the electron density:
\begin{eqnarray}
\delta n_1(\mu)=-\frac{g|\mu|}{2\pi v_\mathrm{F}^2}\Sigma^{(1)}_{|\mu|/v_\mathrm{F},\sgn(\mu)},\label{n_1st2}
\end{eqnarray}
where
\begin{eqnarray}
\Sigma^{(1)}_{\mathbf{p}\gamma}=-\frac1S\sum_{\mathbf{p}'\gamma'}V_{\mathbf{p}-\mathbf{p}'}
F_{\mathbf{p}\gamma\mathbf{p}'\gamma'}f(\epsilon_{\mathbf{p}'\gamma'})\label{Sigma_1st}
\end{eqnarray}
is the $T\rightarrow0$ limit of the electron first-order exchange self-energy, depicted in Fig.~\ref{Fig1}(b).

The explicit expressions for $\Sigma^{(1)}_{\mathbf{p}\gamma}$, calculated beyond the logarithmic term \cite{Gonzalez1},
were presented in \cite{Borghi,Rossi,Peres,Hwang,Li_Hwang,Polini2}. This self-energy can be calculated exactly in terms of
generalized hypergeometric functions, but its expansion
\begin{eqnarray}
\Sigma^{(1)}_{|\mu|/v_\mathrm{F},\sgn(\mu)}=\frac{e^2\mu}{2\varepsilon
v_\mathrm{F}}\left\{\frac12\ln\Lambda+\ln2\right.\nonumber\\
\left.-\frac14-\frac{2\mathcal{C}+1}\pi+\frac{\sgn(\mu)}{4\Lambda}\right\}\label{Sigma_1st2}
\end{eqnarray}
in powers of the dimensionless cutoff $\Lambda=v_\mathrm{F}p_\mathrm{c}/|\mu|$ ($p_\mathrm{c}$ is the cutoff momentum
in the valence band, $\mathcal{C}\approx0.916$ is Catalan's constant) up to $\Lambda^{-1}$ is sufficiently accurate in
the range $6<\Lambda<\infty$, corresponding to the density range $0<|n|<10^{14}\,\mbox{cm}^{-2}$. Thus (\ref{n_1st2})
and (\ref{Sigma_1st2}) allow to determine $\Omega$ and its derivatives in FOA.

\begin{figure}[t]
\begin{center}
\resizebox{1.\columnwidth}{!}{\includegraphics{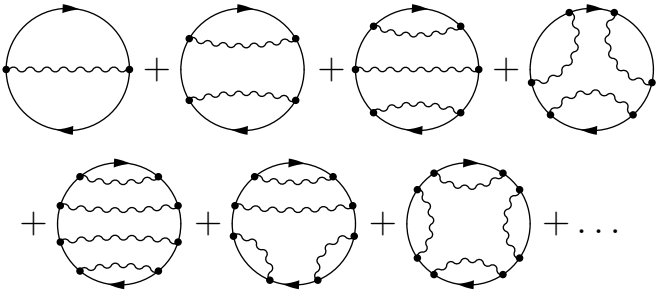}}
\end{center}
\caption{\label{Fig2}The closed connected diagrams for $\delta\Omega$ in the Hartree-Fock approximation.}
\end{figure}

The Hartree-Fock approximation, providing more accurate results than FOA, can be obtained by ``dressing'' the electron
Green functions in Fig.~\ref{Fig1}(a) with exchange self-energy parts. Unfortunately, as is known \cite{Abrikosov}, one
cannot simply replace any bare Green function in closed diagrams by a Hartree-Fock one, because this would result in
overcounting of the diagrams. In fact, in order to obtain HFA starting from the grand canonical ensemble, we need to
calculate the infinite series of diagrams, depicted in Fig.~\ref{Fig2}, with each diagram having a numerical prefactor,
dependent on its symmetry.\footnote{Note that naive calculation of all of these diagrams except the first-order one
directly at $T=0$ will result in zero. Careful evaluation of these ``anomalous'' \cite{Luttinger} or ``rainbow''
\cite{Barnes} diagrams should imply taking the $T\rightarrow0$ limit only after calculating all of them at $T\neq0$ and
summing the full diagrammatic series.}

To overcome this difficulty, we can calculate $\Omega$ by means of the Luttinger-Ward functional \cite{Luttinger},
where all excess diagrams, appearing in the ``overcounted'' thermodynamic potential, are exactly compensated by a
simple expression. In this way, choosing the Hartree-Fock skeleton diagrams (Fig.~\ref{Fig1}), we can calculate
(similarly to Appendix A in \cite{Luttinger2}) the thermodynamic potential in the $T\rightarrow0$ limit:
\begin{eqnarray}
\Omega_\mathrm{HF}=g\sum_{\mathbf{p}\gamma}\left(\epsilon_{\mathbf{p}\gamma}+
\frac{\Sigma_{\mathbf{p}\gamma}^\mathrm{(HF)}}2-\mu\right)
f(\epsilon_{\mathbf{p}\gamma}+\Sigma_{\mathbf{p}\gamma}^\mathrm{(HF)}),\label{Omega_HF1}
\end{eqnarray}
where the Hartree-Fock self-energy is
\begin{eqnarray}
\Sigma_{\mathbf{p}\gamma}^\mathrm{(HF)}=-\frac1S\sum_{\mathbf{p}'\gamma'}
V_{\mathbf{p}-\mathbf{p}'}F_{\mathbf{p}\gamma\mathbf{p}'\gamma'}
f(\epsilon_{\mathbf{p}'\gamma'}+\Sigma_{\mathbf{p}'\gamma'}^\mathrm{(HF)}).\label{Sigma_HF}
\end{eqnarray}

The occupation numbers $f(\epsilon_{\mathbf{p}\gamma}+\Sigma_{\mathbf{p}\gamma}^\mathrm{(HF)})$, entering into these
equations, drop from 1 to 0 at the Fermi surface, where $p=p_\mathrm{F}$, $\gamma=\sgn(\mu)$. Applying (\ref{n}) to
(\ref{Omega_HF1}) and subtracting the background electron density, we obtain the electron density in HFA:
\begin{eqnarray}
n_\mathrm{HF}(\mu)=\frac{g}S\sum_{\mathbf{p}\gamma}\left[
f(\epsilon_{\mathbf{p}\gamma}+\Sigma_{\mathbf{p}\gamma}^\mathrm{(HF)})-\Theta(-\epsilon_{\mathbf{p}\gamma})\right].
\label{n_HF}
\end{eqnarray}
In fact, the expressions (\ref{Omega_HF1})--(\ref{n_HF}) depend on $p_\mathrm{F}$, rather than on $\mu$, therefore it
is more convenient to find the Fermi momentum $p_\mathrm{F}$ from the equation
\begin{eqnarray}
\mu=\epsilon_{p_\mathrm{F},\sgn(n)}+\Sigma_{p_\mathrm{F},\sgn(n)}^\mathrm{(1)}\label{mu_HF}
\end{eqnarray}
and then use Eq.~(\ref{n_HF}), rewritten in the form:
\begin{eqnarray}
n_\mathrm{HF}(\mu)=\sgn(\mu)\frac{gp_\mathrm{F}^2}{4\pi}\label{n_HF1}.
\end{eqnarray}
Here we used the equality $\Sigma^\mathrm{(HF)}_{p_\mathrm{F},\sgn(n)}=\Sigma^{(1)}_{p_\mathrm{F},\sgn(n)}$, following
from (\ref{Sigma_1st}), (\ref{Sigma_HF}) and (\ref{mu_HF}). Solving (\ref{mu_HF})--(\ref{n_HF1}) and integrating
$n_\mathrm{HF}(\mu)$ according to (\ref{n}), we can restore the thermodynamic potential in HFA.

The calculations in (\ref{Omega_HF1})--(\ref{n_HF1}) may appear rather formal, especially in the light of similarity of
(\ref{n0}) and (\ref{n_HF1}), but they demonstrate the essential difference between the first-order and Hartree-Fock
approximations: the latter is \emph{self-consistent}, which means that it actually takes into account an infinite
series of Feynman diagrams (Fig.~\ref{Fig2}) and deals with the renormalized electron dispersion
$\epsilon_{\mathbf{p}\gamma}+\Sigma_{\mathbf{p}\gamma}^\mathrm{(HF)}$ instead of $\epsilon_{\mathbf{p}\gamma}$.

\begin{figure}[t]
\begin{center}
\resizebox{0.9\columnwidth}{!}{\includegraphics{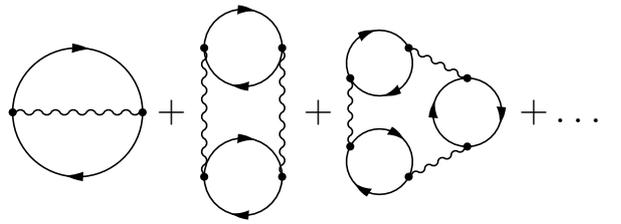}}
\end{center}
\caption{\label{Fig3}The closed connected diagrams for $\delta\Omega$ in the random phase approximation.}
\end{figure}

The random phase approximation for calculating $\Omega$ becomes asymptotically exact in the limit $g\rightarrow\infty$
of large electron state degeneracy \cite{Kotov2,Apenko}. It was also argued that RPA dominates in graphene because of
taking into account all diagrams with infrared divergences \cite{Das_Sarma2}. Recently applicability of RPA has been
confirmed by quick convergence of expansion in RPA-screened interaction \cite{Hofmann}. The sum of diagrams for
$\delta\Omega$ in this approximation, shown in Fig.~\ref{Fig3}, is (see also \cite{Barlas}):
\begin{eqnarray}
\delta\Omega_\mathrm{RPA}=\frac12\sum_{\mathbf{q}}\left\{T\sum_{\omega_k}
\ln\left[1-V_\mathbf{q}\Pi_{\mathbf{q}}(i\omega_k)\right]-nV_\mathbf{q}\right\},\label{Omega_RPA}
\end{eqnarray}
where $\omega_k=2\pi Tk$ are bosonic Matsubara frequencies. The polarizability of the electron gas in graphene
$\Pi_{\mathbf{q}}(\omega)$ was calculated explicitly elsewhere at real \cite{Wunsch,Hwang2} and imaginary \cite{Barlas}
frequencies.

\begin{figure}[t]
\begin{center}
\resizebox{0.85\columnwidth}{!}{\includegraphics{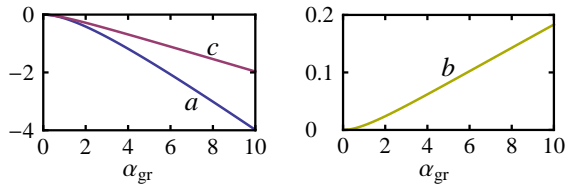}}
\end{center}
\caption{\label{Fig4}The functions $a(\alpha_\mathrm{gr})$, $b(\alpha_\mathrm{gr})$ and $c(\alpha_\mathrm{gr})$ in the
expansion (\ref{Omega_corr}) of the correlation part of $\Omega$ in RPA.}
\end{figure}

It is useful to separate (\ref{Omega_RPA}) into the first-order exchange part (\ref{Omega_1st}) and correlation part
$\delta\Omega_\mathrm{corr}$. In order to obtain analytical results, we expand $\delta\Omega_\mathrm{corr}$ in powers
of $1/\Lambda$ up to $\Lambda^{-2}$:
\begin{eqnarray}
\delta\Omega_\mathrm{corr}=\frac{S|\mu|^3}{4\pi^2v_\mathrm{F}^2}\left\{a(\alpha_\mathrm{gr})
\ln\Lambda+c(\alpha_\mathrm{gr})+\frac{b(\alpha_\mathrm{gr})}{\Lambda^2}\right\},\label{Omega_corr}
\end{eqnarray}
where $\alpha_\mathrm{gr}=ge^2/\varepsilon v_\mathrm{F}$. The functions $a(\alpha_\mathrm{gr})$,
$b(\alpha_\mathrm{gr})$ and $c(\alpha_\mathrm{gr})$, being smooth, can be easily tabulated and approximated in the
physically accessible range $0\leq\alpha_\mathrm{gr}\lesssim10$ (see Fig.~\ref{Fig4}). Our results for
$\delta\Omega_\mathrm{corr}$ are close to those given in \cite{Barlas,Polini,Polini2}.

From (\ref{n}) and (\ref{Omega_corr}) we get the correlation correction
\begin{eqnarray}
\delta n_\mathrm{corr}=-\sgn(\mu)\frac{\mu^2}{4\pi^2v_\mathrm{F}^2}\left\{3a(\alpha_\mathrm{gr})
\ln\Lambda\vphantom{\frac{5b(\alpha_\mathrm{gr})}{\Lambda^2}}\right.\nonumber\\
\left.+3c(\alpha_\mathrm{gr})-a(\alpha_\mathrm{gr})+ \frac{5b(\alpha_\mathrm{gr})}{\Lambda^2}\right\}
\end{eqnarray}
to the electron density $n=n_0+\delta n_1+\delta n_\mathrm{corr}$ in RPA.

Note that the presented RPA is not self-consistent. The Luttinger-Ward theorem can be used to find $\Omega$ in the
self-consistent RPA, which involves all diagrams of Fig.~\ref{Fig3} with the diagrammatic series from Fig.~\ref{Fig2}
inside each loop, i.e. all diagrams without vertex corrections. All calculated quantities (Green functions, self-energy
parts, thermodynamic potential) are consistent among themselves in this approximation. For example, calculation of the
electron density at given $\mu$ through (\ref{n}) or by solving the off-shell Dyson equation
$\mu=\epsilon_{p_\mathrm{F},\sgn(n)}+\Sigma_{p_\mathrm{F},\sgn(n)}(\mu)$ (see \cite{Das_Sarma3}) will yield, unlike
usual RPA, the same results when RPA is treated self-consistently. However, solving a complicated integral equation for
the self-energy is required in the this case.

\section{Analysis of experimental data}\label{sect2}

The basic effect of electron interactions, considered in this article, is a deviation of the dependence $n(\mu)$ from its
noninteracting form (\ref{n0}). This effect can be analyzed in terms of the renormalized Fermi velocity
\begin{eqnarray}
v_\mathrm{F}^*=|\mu|/{p_\mathrm{F}},\label{v_F_ren}
\end{eqnarray}
deviating from $v_\mathrm{F}$ in presence of interactions, or in terms of quantum capacitance $C_\mathrm{Q}$, which is
proportional to $dn/d\mu$. In this article, we use (see also \cite{Li,Yu}) the term ``renormalized Fermi velocity'' in
the meaning of the thermodynamic Fermi velocity (\ref{v_F_ren}), though sometimes this term is referred to the group
velocity of quasiparticles on the Fermi surface, as discussed in \cite{Hwang}.

Our calculations depend on two parameters: the bare Fermi velocity $v_\mathrm{F}$ and the environmental dielectric
constant $\varepsilon$. To estimate them, we analyze experimental data on measured $C_\mathrm{Q}$ or $v_\mathrm{F}^*$
within HFA and RPA, described in the previous section. For our analysis, we use the data from four recent experimental
works \cite{Kretinin,Yu,Chae,Chen2}, where the measured $v_\mathrm{F}^*$ \cite{Yu,Chae} or $C_\mathrm{Q}$
\cite{Kretinin,Chen2} were reported.

In all our calculations, we use, following \cite{Sheehy,Asgari}, the cutoff momentum
$p_\mathrm{c}=1.095\,\mbox{\AA}^{-1}$, found by equating the density of valence band electrons $2/S_0$ to
$gp_\mathrm{c}^2/4\pi$, where $S_0=5.24\,\mbox{\AA}^2$ is the area of graphene elementary cell.

We employ the following fitting procedures: first, we take the actual values of $\varepsilon$, determined by the
substrate material in the experimental setup of each analyzed work \cite{Kretinin,Yu,Chae,Chen2}, and obtain
$v_\mathrm{F}$ through the least square fittings of the measured $v_\mathrm{F}^*$ or $C_\mathrm{Q}$ with RPA
theoretical formulas. Assuming that RPA appropriately takes into account both exchange and correlation effects (see
also \cite{Barnes,Hofmann}), the resulting values of $v_\mathrm{F}$ are expected to be generally adequate.

Second, we take these values of $v_\mathrm{F}$ and fit the same experimental data in HFA, obtaining new effective
values of $\varepsilon$. These quantities turn out to be systematically larger, than the actual material values of
$\varepsilon$, because the screening of the Coulomb interaction, present in RPA and now absent in HFA, should be
mimicked by a stronger environmental screening.

\begin{figure}[t]
\begin{center}
\resizebox{1.0\columnwidth}{!}{\includegraphics{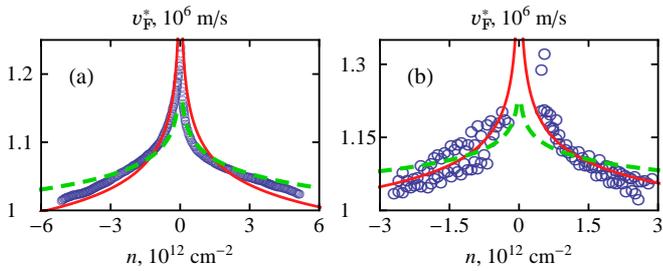}}
\end{center}
\caption{\label{Fig5}Experimental data (blue circles) from \cite{Yu} (a) and \cite{Chae} (b) on renormalized Fermi
velocity $v_\mathrm{F}^*$ as a function of electron density $n$, fitted in the Hartree-Fock (dashed line) and random
phase (solid line) approximations. The data from \cite{Yu} are recalculated with the additional electron density.}
\end{figure}

\begin{figure}[b]
\begin{center}
\resizebox{1.0\columnwidth}{!}{\includegraphics{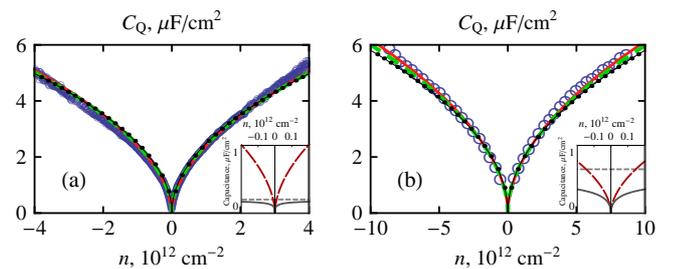}}
\end{center}
\caption{\label{Fig6}Experimental data (blue circles) from \cite{Kretinin} (a) and \cite{Chen2} (b) on quantum
capacitance of graphene $C_\mathrm{Q}$ as a function of electron density $n$, fitted in the noninteracting model
(dotted line) and in the Hartree-Fock (dashed line) and random phase (solid line) approximations. Insets: classical
(dashed line), quantum (broken line, RPA fit) and total (solid line) capacitances per unit area versus electron density
$n$.}
\end{figure}

The parameters $v_\mathrm{F}$ and $\varepsilon$, resulting from our fittings, are collected in Table~\ref{Table}
together with the authors' own estimates. The corresponding experimental points and theoretical curves are shown in
Figs.~\ref{Fig5} and \ref{Fig6}. The comments on each of the considered experimental papers
\cite{Kretinin,Yu,Chae,Chen2}, followed by a short discussion, are given below.

\begin{table}[b]
\centering
\begin{tabular}{|l||c|c||c|c||c|c|}
\hline%
Experiment & \multicolumn{2}{|c||}{Authors' fit} & \multicolumn{2}{|c||}{RPA fit} & \multicolumn{2}{|c|}{HFA fit} \\
\cline{2-7}%
& $v_\mathrm{F}$ & $\varepsilon$ & $v_\mathrm{F}$ & $\varepsilon$ & $v_\mathrm{F}$ & $\varepsilon$ \\
\hline%
Yu et al. \cite{Yu} & 0.850 & 8 & 0.892 & 4.5 & 0.892 & 9.01 \\
\hline%
Chae et al. \cite{Chae} & 0.957 & 3.15 & 0.910 & 3.15 & 0.910 & 8.45 \\
\hline%
Kretinin et al. \cite{Kretinin} & 1 & --- & 1.039 & 4.5 & 1.039 & 14.04 \\
\hline%
Chen et al. \cite{Chen2} & 0.957 & 4.14 & 1.386 & 4.14 & 1.386 & 9.07 \\
\hline%
\end{tabular}
\caption{\label{Table}Fitting parameters for the experimental data on quantum capacitance and renormalized Fermi
velocity, determined by the authors of the corresponding papers and found in our study in HFA and RPA. The RPA values
of $\varepsilon$ are taken according to experimental conditions, and $v_\mathrm{F}$ is given in the units
$10^6\,\mbox{m/s}$.}
\end{table}

In the work \cite{Yu} the capacitance $C$ between AuTi gate electrode and graphene flake, encapsulated in hexagonal
boron nitride (hBN), was accurately measured as a function of the gate voltage $V_\mathrm{g}$. Then $C(V_\mathrm{g})$
was integrated to obtain the electron density:
\begin{eqnarray}
n(V_\mathrm{g})=\frac1{eS}\int\limits_0^{V_\mathrm{g}}C(V_\mathrm{g}')\:dV_\mathrm{g}'.\label{integraton}
\end{eqnarray}
The independently determined classical (or geometrical) capacitance per unit area $C_\mathrm{C}$ allowed then to obtain
the chemical potential $\mu=eV_\mathrm{g}-e^2n/C_\mathrm{C}$ and hence $v_\mathrm{F}^*$ (\ref{v_F_ren}). Using the
first-order renormalization group result \cite{Gonzalez1}
\begin{eqnarray}
\frac{v_\mathrm{F}^*}{v_\mathrm{F}}=1+\frac{\alpha_\mathrm{gr}}8\ln\frac{n_\mathrm{c}}{|n|},\label{v_F_ren1}
\end{eqnarray}
the effective background dielectric constant $\varepsilon=8$ was obtained in \cite{Yu} (the assumed cutoff density
$n_\mathrm{c}=10^{15}\,\mbox{cm}^{-2}$ corresponds to $p_\mathrm{c}=0.56\,\mbox{\AA}^{-1}$). This $\varepsilon$ is much
larger than the actual dielectric constant $\varepsilon=4.5$ of hBN, as expected in the first-order approximation,
neglecting the screening.

The data on $v_\mathrm{F}^*(n)$, presented in \cite{Yu}, demonstrate strong asymmetry and anomalous behavior near CNP,
which are not commented by the authors. We suggest a possible explanation of this anomaly that some nonzero charge is
present on graphene even at zero voltage due to impurities or parasitic external voltage. This excess charge appears as
an integration constant in the right hand side of (\ref{integraton}).

Assuming the additional charge density, equivalent to the electron density $\Delta n=-1.5\times10^9\,\mbox{cm}^{-2}$
and recalculating the dependence $v_\mathrm{F}^*(n)$, we managed to improve substantially agreement between the
experiment \cite{Yu} and our theoretical curves, as shown in Fig.~\ref{Fig5}(a). The both approximations reproduce the
experimental points fairly well.

In \cite{Chae} a graphene sample, placed on a hBN layer on top of oxidized silicon, was studied with the scanning
tunneling spectroscopy in magnetic field. The electron and hole renormalized Fermi velocities were extracted from
Landau level energies at different points of the sample, chosen inside electron and hole puddles. The authors fit the
dependence $v_\mathrm{F}^*(n)$ by an approximate RPA formula using the effective background dielectric constant
$\varepsilon=(1+5.3)/2\approx3.15$, which originates as a half-sum of the dielectric constants of air and
hBN-$\mathrm{SiO}_2$ substrate layer. Our fittings of the data from \cite{Chae}, based on this dielectric constant, are
shown in Fig.~\ref{Fig5}(b).

The work \cite{Kretinin} is focused mainly on electron-hole asymmetry, however measurements of quantum capacitance were
carried out there on high-quality graphene samples in the experimental setup, very similar to that in \cite{Yu}. The
results of the fitting, based on the hBN dielectric constant $\varepsilon=4.5$, are shown in Fig.~\ref{Fig6}(a).
Generally a quantum correction to the classical capacitance in this case is rather small, except the immediate vicinity
of CNP, as shown in the inset.

In the work \cite{Chen2} the inverse compressibility $\kappa^{-1}$ was measured as a function of electron density in
graphene samples on $\mathrm{SiO}_2$ substrate, covered by $\mathrm{Y}_2\mathrm{O}_3$ insulating layer. The data at
$|n|<0.3\times10^{12}\,\mbox{cm}^{-2}$ were excluded from our analysis because of distorting effects of disorder,
appreciable at these concentrations.

The authors of \cite{Chen2} adopt $v_\mathrm{F}=0.957\times10^6\,\mbox{m/s}$ from \cite{Chae} and use the effective
dielectric constant $\varepsilon=(3.9+4.38)=4.14$ to reproduce the measured $\kappa^{-1}$ with the first-order
expression, similar to (\ref{v_F_ren1}). We replotted the data on $\kappa$ in terms of quantum capacitance
$C_\mathrm{Q}$ (see (\ref{compressibility})) and show our fits, based on $\varepsilon=4.14$, in Fig.~\ref{Fig6}(b). As
demonstrated in the inset, a quantum correction to the classical capacitance is significant.

Looking at Table~\ref{Table} and comparing the values of $\varepsilon$, taken according to the experimental conditions
\cite{Yu,Chae,Kretinin,Chen2} and then used in RPA fit, with those obtained via HFA fit, we see, as discussed above,
that in the latter case $\varepsilon$ is larger by 4.5--5.5 (except the case of \cite{Kretinin}, where it is larger by
9.5 by unknown reason). This difference, however, exceeds that following from the simple estimate \cite{Yu}
$\varepsilon_\mathrm{eff}=\varepsilon+\pi ge^2/8v_\mathrm{F}\approx\varepsilon+3.46$, based on a static interband
screening \cite{Hwang2}.

One can also note the anomalously high values of $v_\mathrm{F}$, obtained by fitting the data from \cite{Chen2}. Even
within the authors' theoretical model, the best agreement with the experimental data is achieved at
$v_\mathrm{F}=1.115\times10^6\,\mbox{m/s}$, while the estimate $v_\mathrm{F}=0.957\times10^6\,\mbox{m/s}$, assumed in
\cite{Chen2}, provides the values of $\kappa^{-1}$, which are smaller than the experimental ones. Perhaps the source of
this anomaly is underestimated classical capacitance, used to extract the compressibility from total capacitance.

Lastly, quantum capacitance $C_\mathrm{Q}$, in contrast to $v_\mathrm{F}^*$, does not qualitatively change its
dependence on $n$ when interactions are switched on (see Fig.~\ref{Fig8}(a) in the next section). As a consequence, the
experimental points on $C_\mathrm{Q}$ can be well described by the noninteracting dependence
$C_\mathrm{Q}=e^2\sqrt{g|n|/\pi v_\mathrm{F}^2}$ with $v_\mathrm{F}=1.104\times10^6\,\mbox{m/s}$, $\varepsilon=4.5$ for
the data from \cite{Kretinin} and $v_\mathrm{F}=1.496\times10^6\,\mbox{m/s}$, $\varepsilon=4.14$ for the data from
\cite{Chen2} (see Fig.~\ref{Fig6}).

\section{Many-body effects of Coulomb interactions on characteristics of the electron gas}\label{sect3}

To calculate the quantum capacitance and the renormalized Fermi velocity of the electron gas in graphene, we choose the
value $v_\mathrm{F}\approx0.9\times10^6\,\mbox{m/s}$ of the bare Fermi velocity, consistent with most of the data in
Table~\ref{Table}. We also take three characteristic values of the background dielectric constant, controlling an
interaction strength: $\varepsilon=1$ (suspended graphene), $\varepsilon=4.5$ (graphene, encapsulated in hBN) and
$\varepsilon=8$ (graphene in a strongly screening environment).

\begin{figure}[t]
\begin{center}
\resizebox{1.0\columnwidth}{!}{\includegraphics{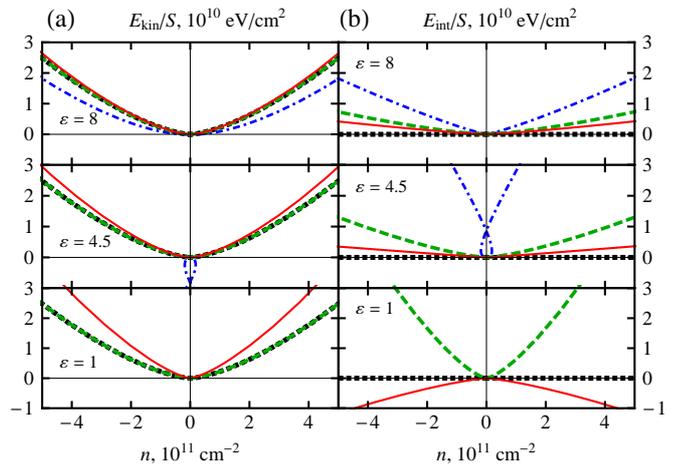}}
\end{center}
\caption{\label{Fig7}Kinetic (a) and interaction (b) energies of the electron gas in graphene, calculated as functions
of electron density $n$ at different dielectric constants $\varepsilon$ in the noninteracting regime (dotted line) and
in the first-order (dash-dotted line), Hartree-Fock (dashed line) and random phase (solid line) approximations.}
\end{figure}

\begin{figure}[t]
\begin{center}
\resizebox{1.\columnwidth}{!}{\includegraphics{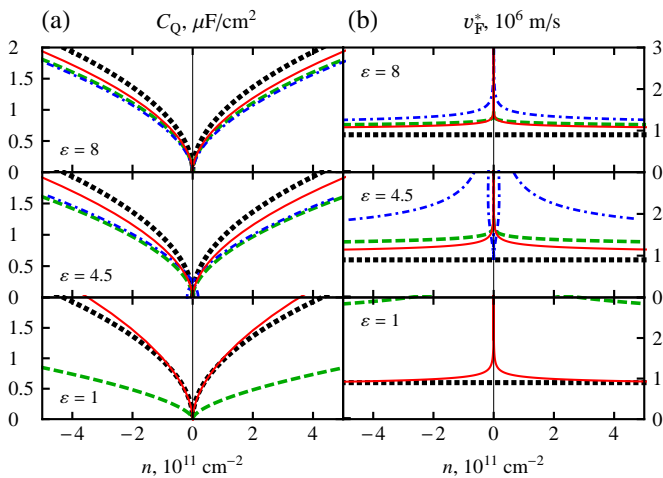}}
\end{center}
\caption{\label{Fig8}Quantum capacitance $C_\mathrm{Q}$ (a) and renormalized Fermi velocity $v_\mathrm{F}^*$ (b),
calculated as functions of electron density $n$ at different dielectric constants $\varepsilon$ in the noninteracting
regime (dotted line) and in the first-order (dash-dotted line), Hartree-Fock (dashed line) and random phase (solid
line) approximations.}
\end{figure}

To get an additional insight into results, we consider the kinetic $E_\mathrm{kin}$ and Coulomb interaction
$E_\mathrm{int}$ energies of the electron gas, which can be found on the basis of the grand canonical ensemble as:
$E_\mathrm{kin}=v_\mathrm{F}(\partial\Omega/\partial v_\mathrm{F})$, $E_\mathrm{int}=e^2(\partial\Omega/\partial e^2)$.
These energies, calculated for the ideal Dirac electron gas and for the interacting gas in different approximations,
are shown in Fig.~\ref{Fig7}. The quantum capacitance and renormalized Fermi velocity, calculated under the same
conditions, are shown in Fig.~\ref{Fig8}. FOA provides reasonable results only in a weak-interacting regime
($\varepsilon\gg1$), thus its results are not shown at $\varepsilon=1$. Even at $\varepsilon=4.5$ it shows such
artifacts as multiple-valuedness of $E_\mathrm{kin}(n)$, $E_\mathrm{int}(n)$, $\mu(n)$ and negative compressibility and
$C_\mathrm{Q}$ near CNP.

In contrast to a usual electron gas with negative exchange energy, the electron exchange self-energy in graphene
(\ref{Sigma_1st2}) is positive due to its chirality \cite{Polini}. Therefore the exchange effects in graphene tend to
increase $v_\mathrm{F}^*$, as seen in Fig.~\ref{Fig8}(a) in FOA and HFA. As a consequence, $C_\mathrm{Q}$ becomes
smaller (Fig.~\ref{Fig8}(b)), because this quantity reflects an effective density of states, which decreases as
$v_\mathrm{F}^*$ increases (note that the interaction-induced change in $C_\mathrm{Q}$ can be essentially diminished by
a proper choice of the fitting parameter $v_\mathrm{F}$ within the non-interacting model, as seen in Fig.~\ref{Fig6}).

For the same reason, the interaction energy $E_\mathrm{int}$, consisting of exchange energies of individual electrons,
is positive in FOA and HFA (Fig.~\ref{Fig7}(b)). The kinetic energy $E_\mathrm{kin}$ decreases in FOA due to decreasing
density (Fig.~\ref{Fig7}(a)). In HFA it does not change in comparison with the non-interacting regime at the same
density, because in the both cases the ground state wave function is the same Slater determinant. Generally HFA
provides more plausible and moderate results than FOA even at $\varepsilon\gg1$, which indicates importance of the
self-consistent treatment of the interactions.

The difference between RPA and FOA results demonstrates correlation effects. As is known, electrons in the correlated
electron liquid tend to be located in average farther from each other in comparison with the mean field picture, thus
the interaction energy decreases. At the same time, the kinetic energy increases because of this additional correlated
motion. The both of these effects are seen in Fig.~\ref{Fig7}. From the other point of view, the correlations partly
compensate the exchange effects \cite{Barlas,Polini2} via screening of the Coulomb interaction. This can be seen in
Fig.~\ref{Fig8} at $\varepsilon=4.5$ and 8, where the RPA curves are situated  between noninteracting and HFA (or FOA)
curves. However, at $\varepsilon=1$ the correlation effects, which are at least quadratic in $\alpha_\mathrm{gr}$, can
even overcompensate the linear in $\alpha_\mathrm{gr}$ exchange effects, resulting in the negative interaction energy
(Fig.~\ref{Fig7}(b)) and increased quantum capacitance (Fig.~\ref{Fig8}(a)).

\begin{figure}[t]
\begin{center}
\resizebox{1.0\columnwidth}{!}{\includegraphics{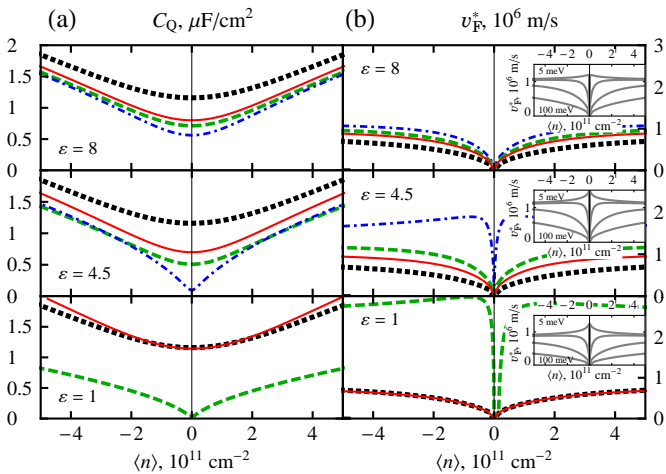}}
\end{center}
\caption{\label{Fig9}The same quantities as in Fig.~\ref{Fig8}, calculated as functions of the average electron density
$\langle n\rangle$ in the Gaussian disorder potential with the spread $s=50\,\mbox{meV}$. Insets in (b):
$v_\mathrm{F}^*$, calculated in RPA as a function of $\langle n\rangle$ with the following values of $s$ (from top to
bottom): 5, 20, 50 and 100 meV.}
\end{figure}

\section{Disorder}\label{sect4}

A random disorder potential $V(\mathbf{r})$, arising in graphene sample due to substrate charge impurities and
corrugations, leads to formation of a spatially varying electron density pattern, emerging as electron and hole puddles
near CNP \cite{Martin,Zhang,Chae,Hajaj,Martin2}. Typical size of the puddles, observed in recent experiments, is
10--20\,nm \cite{Zhang,Chae,Martin2}. Thus the local density approximation, proposed and used
\cite{Hajaj,Kliros,Zebrev,Abergel2,Xu} to calculate the compressibility and quantum capacitance of disordered graphene,
is applicable at the carrier densities $|n|>10^{11}\,\mbox{cm}^{-2}$, when $p_\mathrm{F}^{-1}\lesssim15\,\mbox{nm}$.

In this approximation we assume that the local chemical potential $\mu_\mathrm{loc}(\mathbf{r})$ is established in each
region of a graphene sample in such a way so that the total electrochemical potential
$\mu=V(\mathbf{r})+\mu_\mathrm{loc}(\mathbf{r})$ is constant throughout the sample. Following
\cite{Hajaj,Kliros,Zebrev,Abergel2,Xu,Li2}, we assume the Gaussian distribution of areas of such regions:
\begin{eqnarray}
\rho(V)=\frac1{\sqrt{2\pi}s}e^{-V^2/2s^2}.\label{distr}
\end{eqnarray}
Thus the experimentally observed electron density in graphene sample can be calculated as a spatial average of the
local density $n(\mu_\mathrm{loc})=n(\mu-V)$:
\begin{eqnarray}
\langle n(\mu)\rangle=\int \rho(V)n(\mu-V)\:dV.\label{n_average2}
\end{eqnarray}

The spread $s$ in (\ref{distr}) can be related to the average charge carrier density $|n|$ at $T\to0$, calculated from
(\ref{n_average2})--(\ref{distr}) at CNP: $\langle|n|\rangle=s^2/\pi v_\mathrm{F}^2$. The values of $s$, reported in
the experiments with graphene on $\mathrm{SiO}_2$ \cite{Droscher,Hajaj,Chen1,Xia1,Xu,Zhang,Ponomarenko,Martin,Chen2}
and other substrates \cite{Xia2,Martin2}, or calculated from the corresponding residual carrier densities, range from
$10$ to $130\,\mbox{meV}$. Therefore we assume $s=50\,\mbox{meV}$ to be a typical disorder strength. Similar values are
considered in theoretical works \cite{Kliros,Zebrev,Abergel2,Li2}.

The quantum capacitance and renormalized Fermi velocity, calculated taking into account disorder by replacing $n(\mu)$
in (\ref{compressibility}) and (\ref{v_F_ren}) with $\langle n(\mu)\rangle$, are shown in Fig.~\ref{Fig9}. The major
effect of disorder on $C_\mathrm{Q}$ is its smearing, leading to appearance of a nonzero $C_\mathrm{Q}$ at CNP, where
$C_\mathrm{Q}=0$ in the clean limit (see Fig.~\ref{Fig8}(a)). At the same time, $v_\mathrm{F}^*$ demonstrates quite
unexpected behavior: it falls to zero in the immediate vicinity of CNP. This is due to the fact that the resulting
finite density of states at CNP $\langle D(E=0)\rangle=s/(2^{1/2}\pi^{3/2}v_\mathrm{F}^2)$ implies $\langle
n\rangle\propto\mu$, so that $v_\mathrm{F}^*\propto|\mu|^{1/2}\propto|\langle n\rangle|^{1/2}$. Perhaps this can
explain the anomalous dip of $v_\mathrm{F}^*$ at CNP, observed in \cite{Yu}.

An influence of disorder of various strength on $v_F^*$ is shown in the inset of Fig.~\ref{Fig9}(b). As seen, the peak
near CNP still survives at $s=5\,\mbox{meV}$ and disappears at $s=20\,\mbox{meV}$. According to our estimates, this
disappearance occurs at $s=12$--$15\,\mbox{meV}$ at each value of $\varepsilon$.

\section{Conclusions}\label{concl}

We have considered the many-body effects of Coulomb interactions on such observable quantities of graphene as the
quantum capacitance $C_\mathrm{Q}$, compressibility $\kappa$ and renormalized (thermodynamic) Fermi velocity
$v_\mathrm{F}^*$. Three approximations (FOA, HFA and RPA) are analyzed and applied for massless Dirac electrons.

The recent experimental data on $v_\mathrm{F}^*$ \cite{Yu,Chae} and $C_\mathrm{Q}$ \cite{Kretinin,Chen2} were analyzed
in RPA, with the bare Fermi velocity $v_\mathrm{F}\approx0.9\times10^6\,\mbox{m/s}$ obtained as the result of the least
square fitting. The same experimental data were described by HFA as well, but with larger values of the background
dielectric constant that simulates the screening, absent in this approximation.

Our main conclusions, concerning the influence of Coulomb interactions on $C_\mathrm{Q}$, $v_\mathrm{F}^*$, and kinetic
and interaction energies of the electron gas in graphene, are the following:

a) Kinetic energy increases in presence of the interactions (in RPA) due to correlated motion of electrons.

b) Interaction energy is positive due to the positive exchange energy (as opposed to a usual electron gas)
\cite{Polini}, while it somewhat reduces in RPA due to the correlations, which partly compensate the exchange.

c) The very demonstrative effect of the interactions is the renormalization of $v_\mathrm{F}^*$ to higher values, most
prominent near CNP. In RPA, $v_\mathrm{F}^*$ increases by 50\,\% at lowest achievable carrier densities
$n\sim10^9\,\mbox{cm}^{-2}$ and by 10--20\,\% at moderate densities $n\sim10^{11}\mbox{--}10^{12}\,\mbox{cm}^{-2}$.

d) The quantum capacitance $C_\mathrm{Q}$ decreases in presence of interactions by 10--15\,\% due to effective
reduction of the density of states at higher $v_\mathrm{F}^*$. However, generally it changes only quantitatively,
retaining the same form $C_\mathrm{Q}\propto\sqrt{n}$ as in the noninteracting model. That is why experimentally
measured $C_\mathrm{Q}$ and $\kappa$ are often successfully described in the noninteracting model
\cite{Abergel2,Hajaj,Chen1,Xia1,Ponomarenko,Xu,Zebrev,Martin,Gianazzo}, but with the higher Fermi velocity
$v_\mathrm{F}\approx1.1\times10^6\,\mbox{m/s}$.

The considered theoretical models can be easily generalized to take into account a disorder fluctuating potential in
the local density approximation. Calculations of $C_\mathrm{Q}$ in the model of Gaussian fluctuations with the typical
spread $50\,\mbox{meV}$ show formation of a nonzero $C_\mathrm{Q}$ at CNP, in agreement with experiments. On the
contrary, $v_\mathrm{F}^*$ acquires a dip at CNP, which can even override the logarithmic interaction-induced peak at
disorder potential spread exceeding 12--15 meV. Note should be taken that we expect such disorder-induced dip only in
the thermodynamic Fermi velocity obtained in e.g. quantum capacitance or cyclotron mass measurements, but not in the
quasiparticle Fermi velocity, obtained in measurements of single-particle characteristics.

Finally we can note that studies of graphene quantum capacitance are important for its electronic applications, because
$C_\mathrm{Q}$ dominates in case of ultrathin oxide layer between graphene and a gate (see, e.g.,
\cite{Parrish,Xu2,Zebrev2}). In this case an additional screening by the metallic gate electrode can essentially affect
the many-body corrections to $C_\mathrm{Q}$, as considered in \cite{Asgari}.

The work of A.D.Z. was supported by RFBR, and the work of Y.E.L. and A.A.S. was supported by the HSE Basic Research
Program.

\end{document}